\documentclass[%
twocolumn,%
showpacs,%
showkeys,%
preprintnumbers,%
amsmath,amssymb,%
page%
]{revtex4}
\usepackage{graphicx}
\usepackage{dcolumn}
\usepackage{bm}

\def\timeevol#1{\setbox1=\hbox{$\longmapsto$}\setbox2=\hbox{$
    \scriptstyle #1$}\copy1\kern-.5\wd1\kern-.5\wd2
    \raise-1.2\ht2\copy2\kern-.5\wd2\kern\wd1}

\begin{document}


\title{Continuous Spectra of Generalized Kronig-Penney Model}
\author{
Taksu Cheon$^{a}$
Takaomi Shigehara$^{b}$
}

\affiliation{
${}^{a}$ Laboratory of Physics, Kochi University of Technology,
Tosa Yamada, Kochi 782-8502, Japan \\
${}^{b}$ Department of Information 
Science, 
Saitama University, 
Urawa, Saitama 338-8570, Japan
}

\date{August 16, 2004}

\begin{abstract}
The standard Kronig-Penney model with periodic $\delta$ potentials 
is extended to the cases with generalized contact interactions.  
The eigen equation which determines the dispersion relation 
for one-dimensional periodic array of the generalized contact interactions 
is deduced with the transfer matrix formalism. 
Numerical results are presented which reveal 
unexpected band spectra with broader band gap in higher energy region
for generic model with generalized contact interaction.
\end{abstract}

\pacs{00 mathematical physics, 05 statistical physics}

\maketitle

\section{Introduction} 

The contact interaction occupies a special position
in quantum mechanics.
The system with contact interaction is often rigorously 
solvable \cite{AG88}, 
and also is useful in examining the effect 
of small obstacle on particle motion. 
The first influential work on contact interactions was 
done by Kronig and Penney \cite{KP31}. 
The model, which has potential consisting of a periodic array 
of $\delta$ functions, has been widely regarded as a standard 
reference model in the solid-state physics for more than 
half a century. 

In spite of the seeming simplicity of contact interactions, 
there are several non-trivial aspects which are largely left unexplored. 
Even in the simplest setting of one dimension, 
there has been a historically longstanding problem of  
realizing the generalized contact interaction in the 
small-size limit of a local self-adjoint interaction. 
In one dimension, there are a four-parameter family 
of generalized contact interactions which conserve 
the current at both sides of the interaction \cite{GK85}. 
This  corresponds to the fact that the one-dimensional kinetic energy operator 
$T=-\frac{d^2}{dx^2}$ with domain $C^{\infty}_{0}({\bf R} \backslash\{0\})$ 
has deficiency indices $(2,2)$ \cite{AG88}. 
In non-relativistic formalism, the current operator is given by 
\begin{eqnarray}
\label{eq1-1}
j(x)={\bf \Psi}^{\dagger}(x)\sigma_2{\bf \Psi}(x), 
\end{eqnarray}
where ${\bf \Psi}$ is defined in terms of the wave function $\varphi$ and its 
space derivative as 
\begin{eqnarray}
\label{eq1-2}
{\bf \Psi}(x)= \left( \begin{array}{c} \varphi(x) \\ \frac{1}{2m}
\varphi'(x) \end{array} \right). 
\end{eqnarray}
The matrix $\sigma_2$ in Eq.(\ref{eq1-1}) 
is the second component of Pauli matrices; 
\begin{eqnarray}
\label{eq1-3}
\sigma_2=\left( \begin{array}{cc}
0 & -i \\ i & 0 \end{array} \right). 
\end{eqnarray}
In Eq.(\ref{eq1-2}), $m$ is the mass of a particle. 
If we put a contact interaction at $x=0$, 
the connection of ${\bf \Psi}$ between both sides of the origin
can be characterized by 
\begin{eqnarray}
\label{eq1-4}
{\bf \Psi}(+0) = {\cal V}{\bf \Psi}(-0), 
\end{eqnarray}
where the current conservation $j(+0)=j(-0)$ requires 
\begin{eqnarray}
\label{eq1-5}
{\cal V}^{\dagger} \sigma_2 {\cal V}=\sigma_2. 
\end{eqnarray}
The generic solution of Eq.(\ref{eq1-5}) is given by 
\begin{eqnarray}
\label{eq1-6}
{\cal V} = e^{i\theta} 
\left(\begin{array}{cc}
\gamma & \delta \\ 
\beta & \alpha
      \end{array}
\right),
\end{eqnarray}
where $\theta\in{\bf R}$ and $\alpha\gamma-\beta\delta=1$. 
The condition (\ref{eq1-6}) covers a four-parameter family.
(See \cite{AD98,CF01,TF01} for the discussions on
rigorous characterization of the full
parameter space.)
In addition to 
the usual $\delta$ potential 
\begin{eqnarray}
\label{eq1-7}
{\cal V}_{\delta}(v) =
\left(
\begin{array}{cc}
1 & 0 \\
v & 1 
\end{array}
\right)   ,
\end{eqnarray}
this family includes the
so-called $\varepsilon$ potential 
(also known by a misnomer, $\delta'$ potential).
The $\varepsilon$ potential induces such 
a boundary condition that 
the wave function has continuous first derivative 
on the right and left, but it has a jump 
proportional to the first derivative \cite{S86b}; 
\begin{eqnarray}
\label{eq1-8}
{\cal V}_{\varepsilon}(u) =  
\left(
\begin{array}{cc}
1 & u  \\
0 & 1 
\end{array}
\right).   
\end{eqnarray}
In \cite{CS98a}, we have constructed the 
$\varepsilon$ potential in the small distance limit of 
three nearby $\delta$ potentials and realized 
in terms of usual $\delta$ and $\varepsilon$ potentials 
a three-parameter family of self-adjoint extensions 
under the assumption of time reversal symmetry. 
This treatment have been generalized with finite range
potential with double short-range limit with different scales \cite{ENZ01}.
Also, the relativistic origin of $\varepsilon$ potential 
has been discussed in \cite{SM99a}. 
Contrary to the general belief, 
some of the non-$\delta$ contact
interactions appears in realizable settings in quite a natural
manner \cite{BBM95}, and, in fact, they are essential in understanding 
the nature of one-dimensional many-body system \cite{CS99}.

In light of these facts,
it is quite interesting to examine the nature of one-dimensional
system
with a periodic array of generalized contact interactions.  
For this purpose, we generalize the Kronig-Penney model 
to the cases of generalized point interactions. 
We are particularly interested in the question whether the original 
Kronig-Penney model with $\delta$-interaction is a generic zero-range 
limit of more realistic models with finite-range interactions.
In this paper, we assume that 
the system has time-reversal symmetry. 
In this case, we have $\theta=0$ in Eq.(\ref{eq1-6}), hence 
\begin{eqnarray}
\label{eq1-9}
{\cal V}=\left(\begin{array}{cc}
\gamma & \delta \\ 
\beta & \alpha
               \end{array}
\right)
\in SL(2,{\bf R}).  
\end{eqnarray}
After giving the formulation of the 
generalized Kronig-Penney model 
with generic contact interactions in Sect.2,  
some elementary numerical results are presented in Sect.3. 
It will be shown that 
the standard periodic array 
of $\delta$ potentials has a specific band structure, 
compared to the generic cases;  
The band gap tends to disappear in the 
high energy limit for periodic $\delta$ array,    
while the band width becomes narrower for higher 
bands in generic cases. 
The current work is summarized in Sect.4.  

\section{Formalism}

We consider a one-dimensional periodic array 
of a generalized contact interaction, 
the connection condition of which is described 
by ${\cal V}$ in Eq.(\ref{eq1-9}). 
We assume that the interactions are located  
at $x=na$, ($n=0, \pm 1, \pm 2,\cdots$). 
Here we denote the lattice interval by $a$. 
The assumed potential is shown in Fig.1.  
%
\begin{figure}
\includegraphics[width=6.2cm]{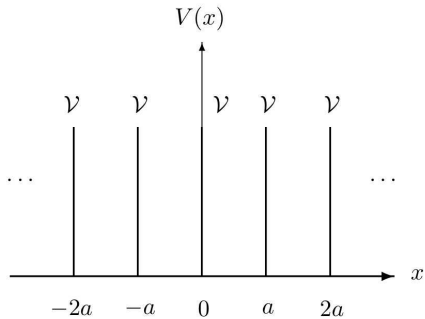}
\caption
{\label{fig1}
Periodic potential of Kronig-Penny model with 
generalized contact interaction. 
The two-by-two matrix ${\cal V}\in SL(2,{\bf R})$ represents 
the connection condition at each interaction in the transfer 
matrix formalism. 
}
\end{figure}
%

\noindent
Schr\"{o}dinger equation is given by 
\begin{eqnarray}
\label{eq2-1}
-\frac{1}{2m}\varphi''(x)=E\varphi(x), \ \ (x\neq na)
\end{eqnarray}
with the boundary condition 
\begin{eqnarray}
\label{eq2-2}
{\bf \Psi}(na+0) = {\cal V}{\bf \Psi}(na-0)  
\end{eqnarray}
at $x=na$. In the transfer matrix formalism, 
Eq.(\ref{eq2-1}) is written as 
the first-order coupled equation; 
\begin{eqnarray}
\label{eq2-3}
{\bf \Psi}'(x) = {\cal H}{\bf \Psi}(x), \ \ 
(x \ne na), 
\end{eqnarray}
where we define 
\begin{eqnarray}
\label{eq2-4}
{\cal H} = \left( \begin{array}{cc}
0   & 2m  \\
-\frac{k_0^2}{2m}  & 0
\end{array} \right)
\end{eqnarray}
with $k_0=\sqrt{2mE}$. 
The solution of Eq.(\ref{eq2-3}) is written as 
\begin{eqnarray}
\label{eq2-4b} 
{\bf \Psi}(x)={\cal G}(x-x_0) {\bf \Psi}(x_0) 
\end{eqnarray}
by using the exponential function of ${\cal H}x$; 
\begin{eqnarray}
\label{eq2-5}
{\cal G}(x)\equiv e^{{\cal H} x} 
= \cos (k_0x) I_2 + \frac{\sin (k_0x)}{k_0}{\cal H}.   
\end{eqnarray}
Here we assume $na \notin [x_0,x]$ in Eq.(\ref{eq2-4b}). 
In Eq.(\ref{eq2-5}), $I_2$ is the two-by-two identity matrix. 
Note that $\det {\cal G}(x)=1$ because of $Tr {\cal H}=0$.  

We can see  
\begin{eqnarray}
\label{eq2-6}
{\cal G}(x){\bf u}_{\pm k_0}=e^{\pm ik_0x}{\bf u}_{\pm k_0}
\end{eqnarray}
with 
\begin{eqnarray}
\label{eq2-7}
{\bf u}_{\pm k_0}=\frac{1}{\sqrt{2}}
\left( \begin{array}{c}
1 \\ \pm i\frac{k_0}{2m}
\end{array} \right). 
\end{eqnarray}
Namely, ${\cal G}(x)$ has eigenvalues 
$e^{\pm ik_0x}$ with the associated eigenfunction ${\bf u}_{\pm k_0}$ 
in Eq.(\ref{eq2-7}). 
Also the complex conjugate of ${\cal G}(x)$ satisfies  
\begin{eqnarray}
\label{eq2-8}
{\cal G}^{\dagger}(x){\bf v}_{\pm k_0}=e^{\mp ik_0x}{\bf v}_{\pm k_0}
\end{eqnarray}
with 
\begin{eqnarray}
\label{eq2-9}
{\bf v}_{\pm k_0}=\frac{1}{\sqrt{2}}
\left( \begin{array}{c}
1 \\ \pm i\frac{2m}{k_0}
\end{array} \right).   
\end{eqnarray}
That is, ${\cal G}^{\dagger}(x)$ has 
eigenvalues $e^{\mp ik_0x}$ with the associated eigenfunction 
${\bf v}_{\pm k_0}$ in Eq.(\ref{eq2-9}). 
The eigenfunctions ${\bf u}_{\pm k_0}$ and ${\bf v}_{\pm k_0}$
satisfy the bi-orthogonal relations; 
\begin{eqnarray}
\label{eq2-10}
{\bf v}_{\pm k_0}^{\dagger}{\bf u}_{\pm k_0}=1, \hspace{3ex}
{\bf v}_{\mp k_0}^{\dagger}{\bf u}_{\pm k_0}=0. 
\end{eqnarray}

By Bloch theorem, we can set 
\begin{eqnarray}
\label{eq2-11}
\varphi(x)=e^{ikx}u(x), \ \ \ \ (k\in{\bf R}) 
\end{eqnarray}
in Eq.(\ref{eq2-1}), where 
the function 
$u$ has period $a$; 
\begin{eqnarray}
\label{eq2-12}
u(x+a)=u(x). 
\end{eqnarray}
Since 
\begin{eqnarray}
\label{eq2-13}
\varphi'(x) & = & e^{ikx}(u'(x)+iku(x)), \\
\label{eq2-14}
\varphi''(x) & = & e^{ikx}(u''(x)+2iku'(x)-k^2u(x)), 
\end{eqnarray}
we obtain the equation for $u$; 
\begin{eqnarray}
\label{eq2-15}
-u''(x)-2iku'(x)+k^2u(x)=k_0^2u(x), 
\end{eqnarray}
($x \neq na$). 
In the vector notation 
\begin{eqnarray}
\label{eq2-16}
{\bf \tilde{\Psi}}(x)= \left( \begin{array}{c} u(x) \\ \frac{1}{2m}
u'(x) \end{array} \right), 
\end{eqnarray}
we rewrite Eq.(\ref{eq2-15}) as 
\begin{eqnarray}
\label{eq2-17}
{\bf \tilde{\Psi}}'(x) = {\tilde{\cal H}}{\bf \tilde{\Psi}}(x), 
\hspace{3ex}
(x\neq na) 
\end{eqnarray}
with 
\begin{eqnarray}
\label{eq2-18}
{\tilde{\cal H}}=
\left( \begin{array}{cc} 0 & 2m \\ 
\frac{k^2-k_0^2}{2m} & -2ik \\ \end{array} \right). 
\end{eqnarray}
It is clear from Eqs.(\ref{eq2-11}) and (\ref{eq2-13}) that 
the two vectors ${\bf \Psi}(x)$ and  ${\bf \tilde{\Psi}}(x)$ 
are related by 
\begin{eqnarray}
\label{eq2-19}
{\bf \Psi}(x)={\cal M}{\bf \tilde{\Psi}}(x)
\end{eqnarray}
with 
\begin{eqnarray}
\label{eq2-20}
{\cal M} = e^{ikx} \left( \begin{array}{cc}
1 & 0 \\ \frac{ik}{2m} & 1 
\end{array} \right). 
\end{eqnarray}
The solution of Eq.(\ref{eq2-17}) is given by 
\begin{eqnarray}
\label{eq2-21}
{\bf \tilde{\Psi}}(x)={\tilde{\cal G}}(x-x_0) {\bf \tilde{\Psi}}(x_0), 
\end{eqnarray}
where ${\tilde{\cal G}}(x)$ is the exponential function of 
${\tilde{\cal H}} x$;  
\begin{eqnarray}
\label{eq2-22}
& &
{\tilde{\cal G}}(x) 
\\ \nonumber
& = & \!\!
  e^{{\tilde{\cal H}} x} 
\\ \nonumber
& = & \!\!
e^{-ikx}
\left[ { \cos(k_0 x) I_2 
+ \frac{\sin(k_0x)}{k_0}
  \left( \begin{array}{cc}
  ik & 2m \\ \frac{k^2-k_0^2}{2m} & -ik \\ 
  \end{array} \right) } 
\right]. 
\end{eqnarray}
Here we assume $na\notin [x_0,x]$. 
Since $Tr {\tilde{\cal H}}=-2ik$, we have 
$\det {\tilde{\cal G}}(x) = e^{(Tr {\tilde{\cal H}}) x}=
e^{-2ikx}$. 
It is seen from Eqs.(\ref{eq2-2}) and (\ref{eq2-19}) that 
the connection condition for ${\bf \tilde{\Psi}}$ at $x=na$ is given by 
\begin{eqnarray}
\label{eq2-23}
{\bf \tilde{\Psi}}(na+0) = {\tilde{\cal V}}{\bf \tilde{\Psi}}(na-0) 
\end{eqnarray}
with
\begin{eqnarray}
\label{eq2-24}
{\tilde{\cal V}} & = & {\cal M}^{-1} {\cal V} {\cal M} \nonumber \\
& = & 
\left( \begin{array}{cc}
1 & 0 \\ -\frac{ik}{2m} & 1 
\end{array} \right) 
{\cal V}
\left( \begin{array}{cc}
1 & 0 \\ \frac{ik}{2m} & 1 
\end{array} \right). 
\end{eqnarray}
Note that $\det {\tilde{\cal V}}=\det{\cal V}=1$. 
The periodicity for $u$ in Eq.(\ref{eq2-12}) is equivalent to 
\begin{eqnarray}
\label{eq2-25}
{\bf \tilde{\Psi}}(x+a)=
{\bf \tilde{\Psi}}(x). 
\end{eqnarray}
In particular, with $x=na-0$, we have 
\begin{eqnarray}
\label{eq2-26}
{\tilde{\cal G}}(a){\tilde{\cal V}}{\bf \tilde{\Psi}}(na-0)=
{\bf \tilde{\Psi}}(na-0),  
\end{eqnarray}
since 
\begin{eqnarray}
\label{eq2-26b}
{\bf \tilde{\Psi}}((n+1)a-0) & = & 
{\tilde{\cal G}}(a){\bf \tilde{\Psi}}(na+0) \nonumber \\  
& = & {\tilde{\cal G}}(a){\tilde{\cal V}}{\bf \tilde{\Psi}}(na-0). 
\end{eqnarray}
Eq.(\ref{eq2-26}) requires the two-by-two matrix 
${\tilde{\cal G}}(a){\tilde{\cal V}}$ has eigenvalue $1$:   
\begin{eqnarray}
\label{eq2-27}
\det( I_2-{\tilde{\cal G}}(a){\tilde{\cal V}} )=0. 
\end{eqnarray}
Since 
$\det ( {\tilde{\cal G}}(a){\tilde{\cal V}} ) =
\det {\tilde{\cal G}}(a)  =e^{-2ika}$, 
the other eigenvalue of the matrix 
${\tilde{\cal G}}(a){\tilde{\cal V}}$ 
should be $e^{-2ika}$.  
This indicates 
\begin{eqnarray}
\label{eq2-28}
Tr ( {\tilde{\cal G}}(a){\tilde{\cal V}} ) =   
1+e^{-2ika}. 
\end{eqnarray}
A simple matrix calculation shows 
\begin{eqnarray}
\label{eq2-29}
M\tilde{{\cal H}}M^{-1}= -ikI_2 + {\cal H},  
\end{eqnarray}
which gives the relation between 
$\tilde{\cal{G}}(x)$ and ${\cal G}(x)$;  
\begin{eqnarray}
\label{eq2-30}
M \tilde{\cal{G}}(x) M^{-1}=e^{M\tilde{{\cal H}}M^{-1}x}
=e^{-ikx}{\cal G}(x). 
\end{eqnarray}
From Eq.(\ref{eq2-30}), we obtain 
\begin{eqnarray}
\label{eq2-31}
{\tilde{\cal G}}(a){\tilde{\cal V}} 
=e^{-ika}{\cal M}^{-1}{\cal G}(a){\cal V}{\cal M}. 
\end{eqnarray}
Thus we conclude that the condition (\ref{eq2-28}) is equivalent to 
\begin{eqnarray}
\label{eq2-32}
Tr ( {\cal G}(a){\cal V} ) 
= 2\cos(ka).   
\end{eqnarray}
Eq.(\ref{eq2-32}) 
determines the dispersion relation for a periodic array 
of the  generalized contact interaction characterized 
by the connection condition (\ref{eq1-9}). 
Inserting Eqs.(\ref{eq1-9}) and (\ref{eq2-5}) into 
Eq.(\ref{eq2-32}), 
we obtain the eigenvalue equation
\begin{eqnarray}
\label{eq2-33}
(\alpha+\gamma)\cos(k_0a)+\sin(k_0a)
\left(\frac{2m}{k_0}\beta-\frac{k_0}{2m}\delta\right) \nonumber \\
= 2\cos (ka). 
\end{eqnarray}

\section{Numerical Examples}

In this section, we give numerical examples of  
the band spectrum in several cases; the usual 
periodic $\delta$, periodic $\varepsilon$ as well as 
periodic array with 
some typical one-parameter families of 
generic contact interactions. 
It will be shown that the band structure of 
periodic $\delta$ potential is {\em not} generic; 
The band width becomes broader even in 
the $k$ (wave number) space for $\delta$ array 
as the energy increases, contrary to the generic 
cases. 
Throughout this section, 
we take the mass of particle $m=1/2$ and the lattice interval 
$a=1$ in numerical calculations. 

Let us begin with the periodic $\delta$ potentials. 
From Eq.(\ref{eq2-33}) together with Eq.(\ref{eq1-7}), 
we obtain the eigenvalue equation 
for $\delta$ array; 
\begin{eqnarray}
\label{eq3-1}
\cos(k_0)+\frac{v\sin(k_0)}{2k_0}=\cos (k), 
\end{eqnarray}
where $v$ is the strength of each $\delta$. 
Eq.(\ref{eq3-1}) is the well-known result 
in the standard Kronig-Penney model. 
Fig.2(a) shows the band spectrum as a function of $v$. 
With the parameters $v$ and $E=k_0^2$ in the shaded region, 
the equation (\ref{eq3-1}) has the solution $k$. 
The condition for the existence of the solution is given by 
\begin{eqnarray}
\label{eq3-2}
\left| \cos(k_0)+\frac{v\sin(k_0)}{2k_0} \right| \leq 1. 
\end{eqnarray}
For $v=0$, one sees the continuous spectrum for positive energy 
and there is no bound states, as expected. 
%
\begin{figure}
\includegraphics[width=6.2cm]{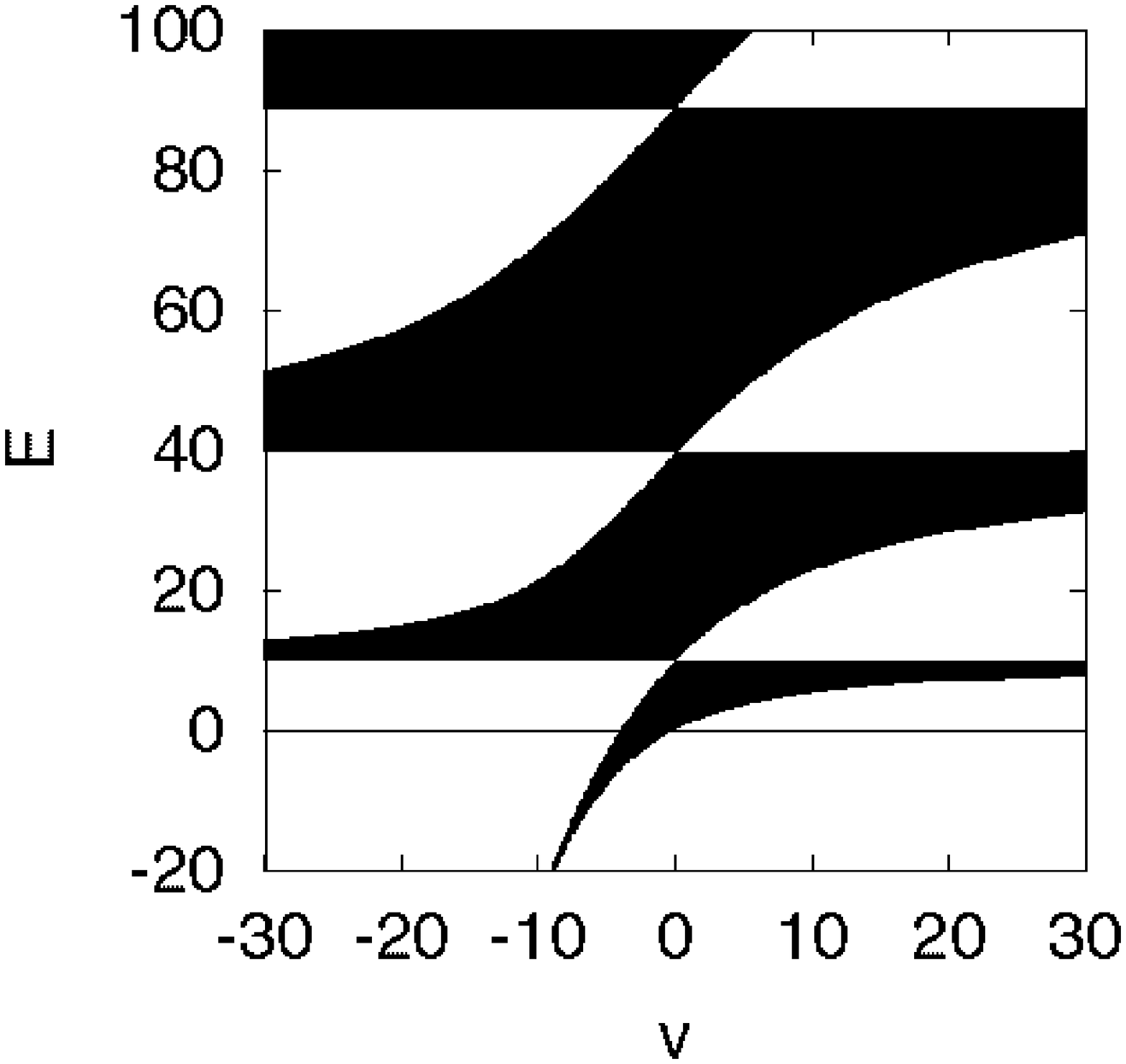}
\includegraphics[width=6.2cm]{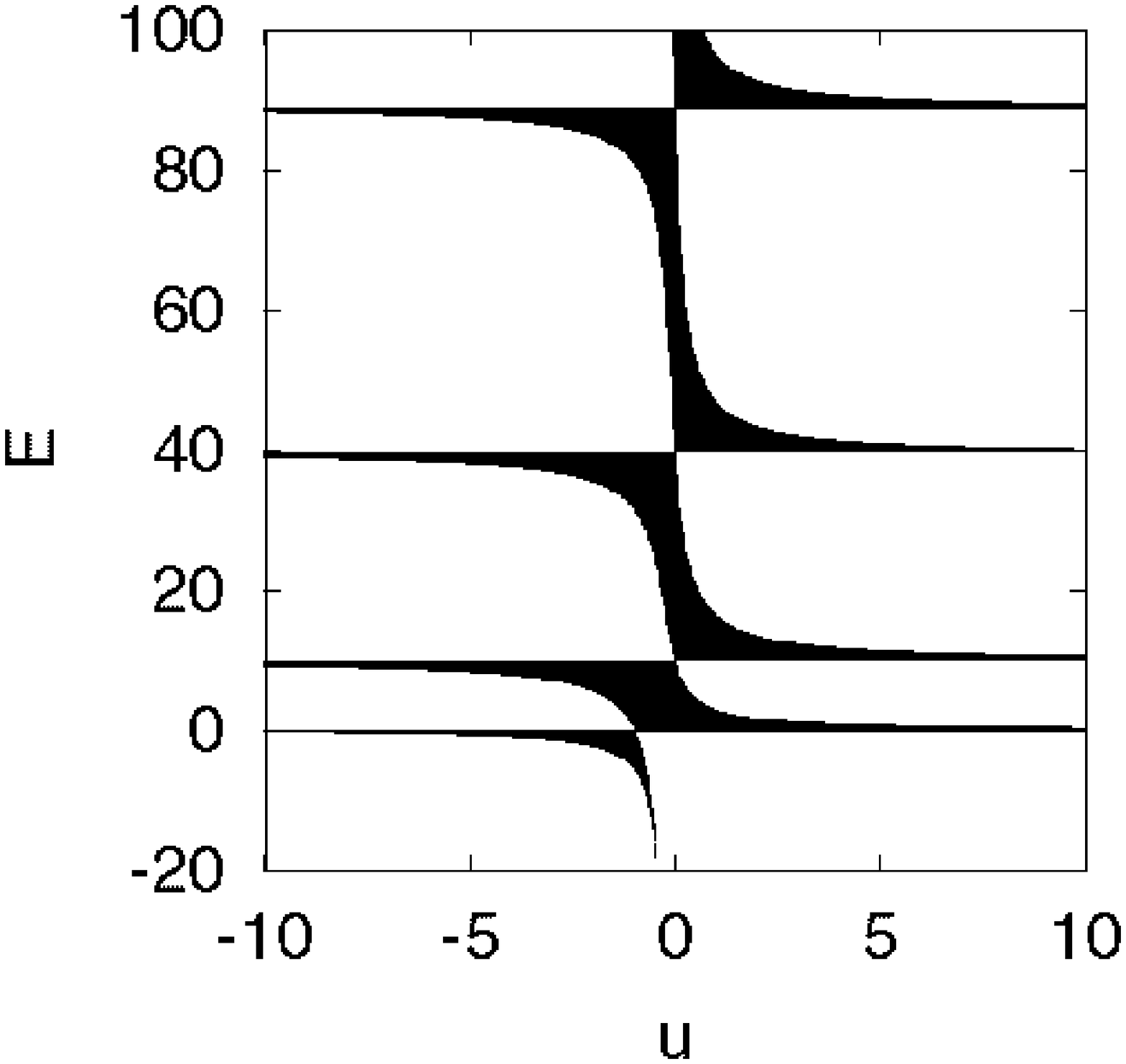}
\includegraphics[width=6.2cm]{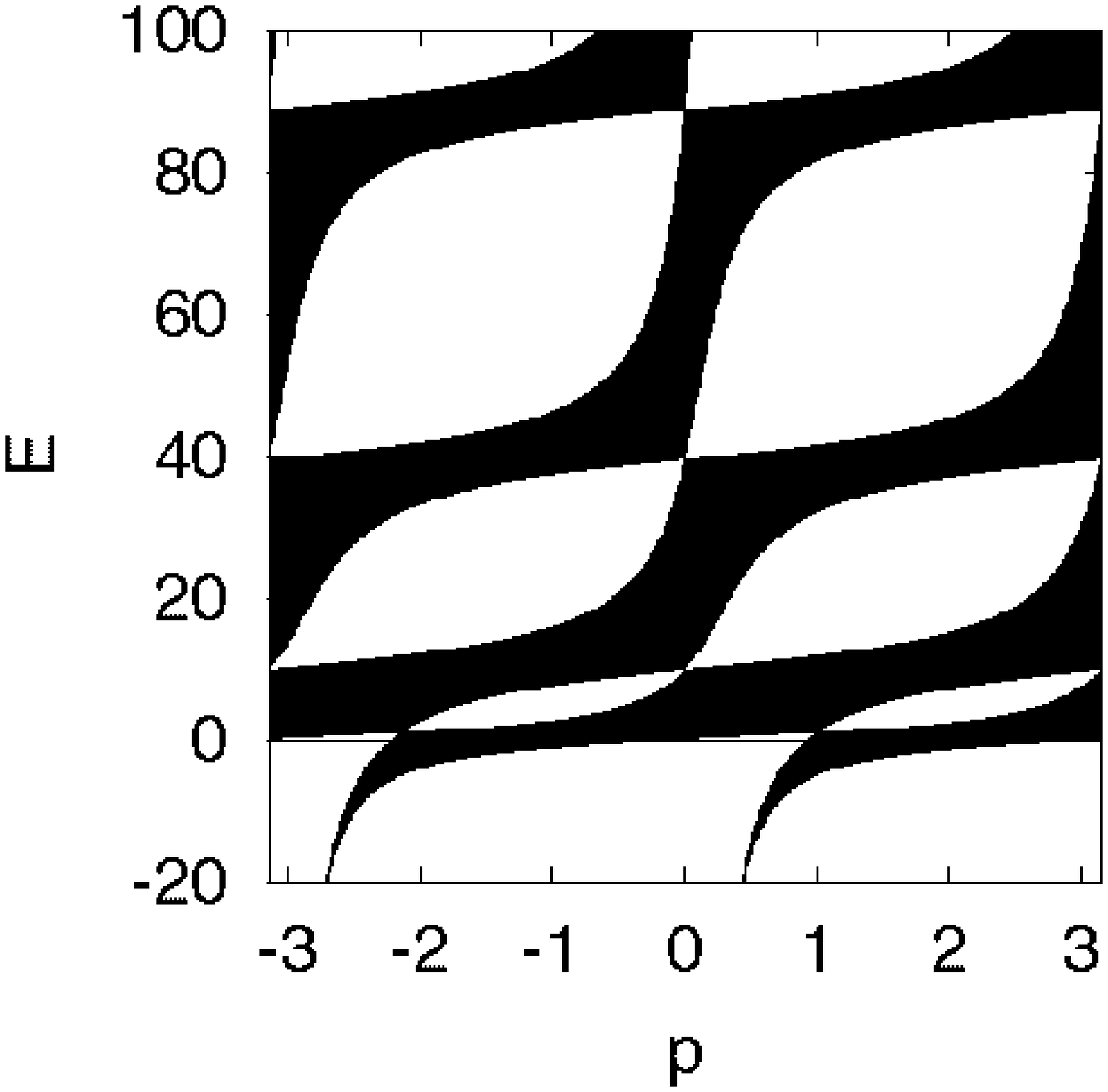}
\includegraphics[width=6.2cm]{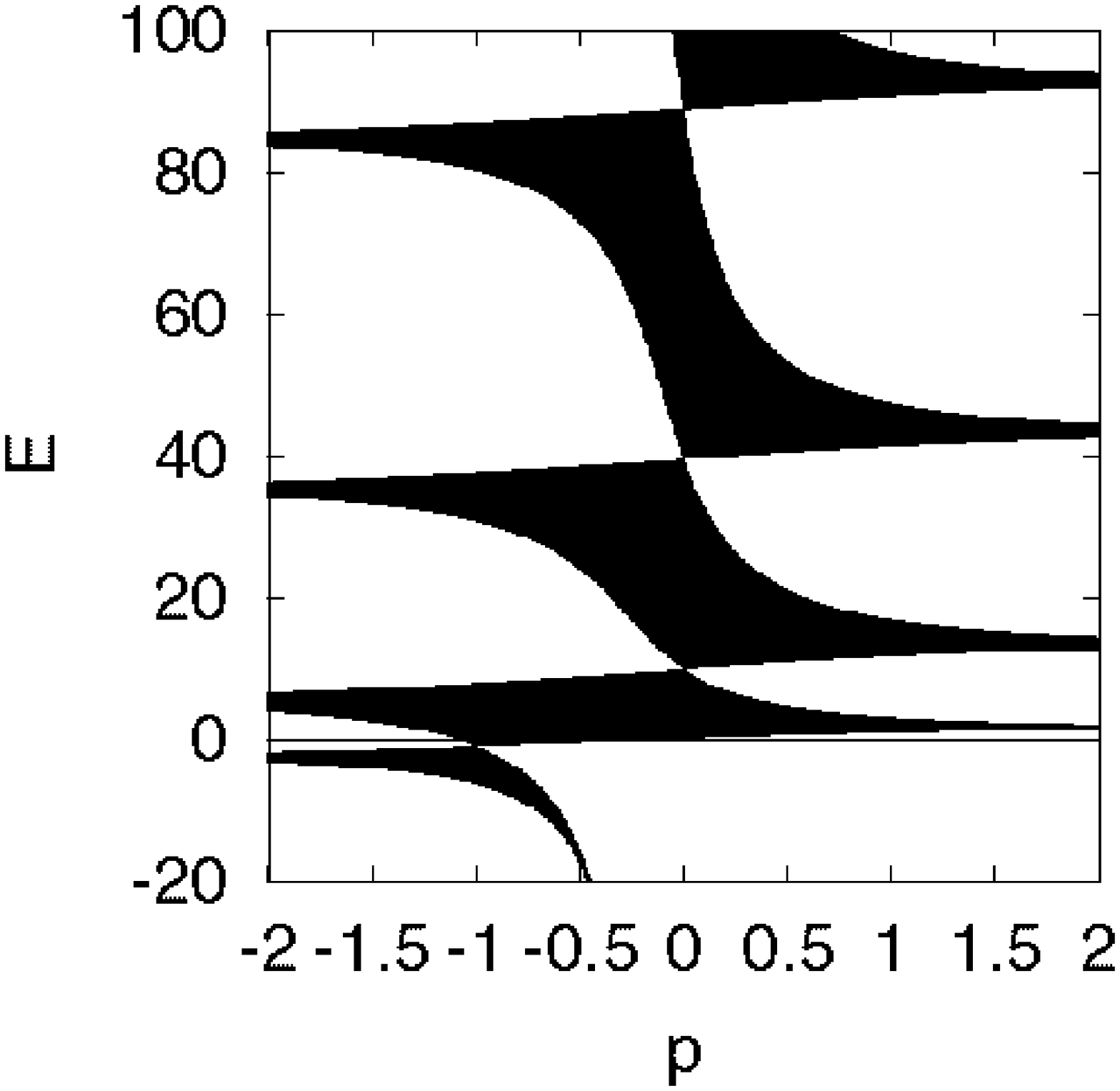}
\caption
{\label{fig2}
Band spectrum for a periodic array of generalized contact 
interaction characterized by one-parameter family of 
connection conditions; 
Eqs.(7), (8), (49), and (51) in (a), (b), (c), and (d), 
respectively. 
In all cases, the horizontal axis is the 
parameter associated with each family,  
while the vertical axis is the energy $E=\frac{k_0^2}{2m}$. 
The mass is set to $m=1/2$. The lattice interval is set to 
$a=1$. 
}
\end{figure}

\vspace{5mm}
\noindent
For $v\neq 0$, the band structure appears and the band width 
becomes narrower as the strength $|v|$ increases. 
In the limit of $|v|\longrightarrow\infty$, one obtains 
point spectrum $E=(n\pi)^2$. 
For negative $v$, there is a negative energy band which 
goes to the minus infinity in the limit of  
$v\longrightarrow-\infty$. 
In both limits $v\longrightarrow \pm\infty$, 
the boundary condition around each $\delta$ 
corresponds to the so-called separated boundary condition. 
Each region between two neighboring $\delta$'s are separated 
from the other regions and 
the wave function satisfies the Dirichlet boundary condition 
such that it vanishes at each $\delta$. 

For $\varepsilon$ potential of strength $u$, 
we obtain the eigenvalue equation 
\begin{eqnarray}
\label{eq3-3}
\cos(k_0)-\frac{uk_0\sin(k_0)}{2}
=\cos (k).
\end{eqnarray}
Fig.2(b) shows the band spectrum as a function of $u$. 
The condition for the existence of the solution 
is given by 
\begin{eqnarray}
\label{eq3-4}
\left| \cos(k_0)-\frac{uk_0\sin(k_0)}{2} \right| \leq 1. 
\end{eqnarray}
The case of $u=0$ corresponds to the free space.  
For $u\neq 0$, the band structure appears and the band width 
becomes rapidly narrower as the strength $|u|$ increases. 
In the limit of $|u|\longrightarrow\infty$, one obtains 
point spectrum $E=(n\pi)^2$. 
In this limit, 
each region between two neighboring $\varepsilon$'s are separated 
from the other regions and 
the wave function satisfies the Neumann boundary condition 
such that its derivative vanishes at each $\varepsilon$. 
For negative $u$, there is a negative energy band which 
goes to the minus infinity in the limit of  
$u\longrightarrow -0$. 

To see the generic cases, 
we show in Fig2.(c) the band spectrum for 
the one-parameter family  
of the connection condition;  
\begin{eqnarray}
\label{eq3-5}
{\cal V} = \left( \begin{array}{cc} 
\cos p & -\sin p \\ \sin p & \cos p 
\end{array} \right), \ \ \ 
(-\pi < p \leq \pi). 
\end{eqnarray}
The band spectrum is determined by 
\begin{eqnarray}
\label{eq3-6}
\hspace*{-2ex}
\left| 
\cos(p)\cos(k_0)+
\frac{\sin(p)\sin(k_0)}{2}\left( k_0 + \frac{1}{k_0}\right) \right|
\leq 1. 
\end{eqnarray}
With $p\neq 0, \pi$, we see the band structure as in other generic cases. 
The band spectrum has period $\pi$, as seen from Eq.(\ref{eq3-6}) and 
the energy spectrum for $p=\pi$ is the same as in the free 
space ($p=0$).  However, the wave function for $p=\pi$ 
differs from the free one. Indeed, it has the same amplitude as 
the continuous wave function in the free space, but it changes 
the phase and as a result it is discontinuous at each obstacle. 
Such ``duality'' of $p=0$ and $p=\pi$ induces double spiral structure 
in energy spectrum \cite{C98}, as implied in Fig.2(c). 

Fig.2(d) shows the case for the 
one-parameter family of 
\begin{eqnarray}
\label{eq3-7}
{\cal V} = \left( \begin{array}{cc} 
\cosh p & \sinh p \\ \sinh p & \cosh p 
\end{array} \right), \ \ \  
(p \in {\bf R}). 
\end{eqnarray}
In this case, the band spectrum is determined by 
\begin{eqnarray}
\label{eq3-8}
\hspace*{-5ex}
\left| \cosh(p)\cos(k_0)-
\frac{\sinh(p)\sin(k_0)}{2}\left( k_0 - \frac{1}{k_0}\right) \right|
\leq 1.  
\end{eqnarray} 
As in other cases, the band width becomes gradually narrower, 
as the perturbation becomes larger. 

An important point is that the $\delta$ potential, among 
generic point interactions,
has rather a special energy dependence 
of band structure. To see this, we show 
the band spectrum including higher energy region 
in Figs.3(a)-(d), which correspond to Figs.2(a)-(d), 
respectively. 
In Fig.3, the vertical axis is the wave number $k_0$, 
instead of the energy $E$, which is suitable for our 
present purpose,  
since the state density is constant [of $O(k^0)$] in one dimension.  
One can see that in generic cases including $\varepsilon$, 
the band width becomes narrower for higher bands, while 
the band gap tends to disappear for $\delta$ array. 
%

\begin{figure}
\includegraphics[width=6.2cm]{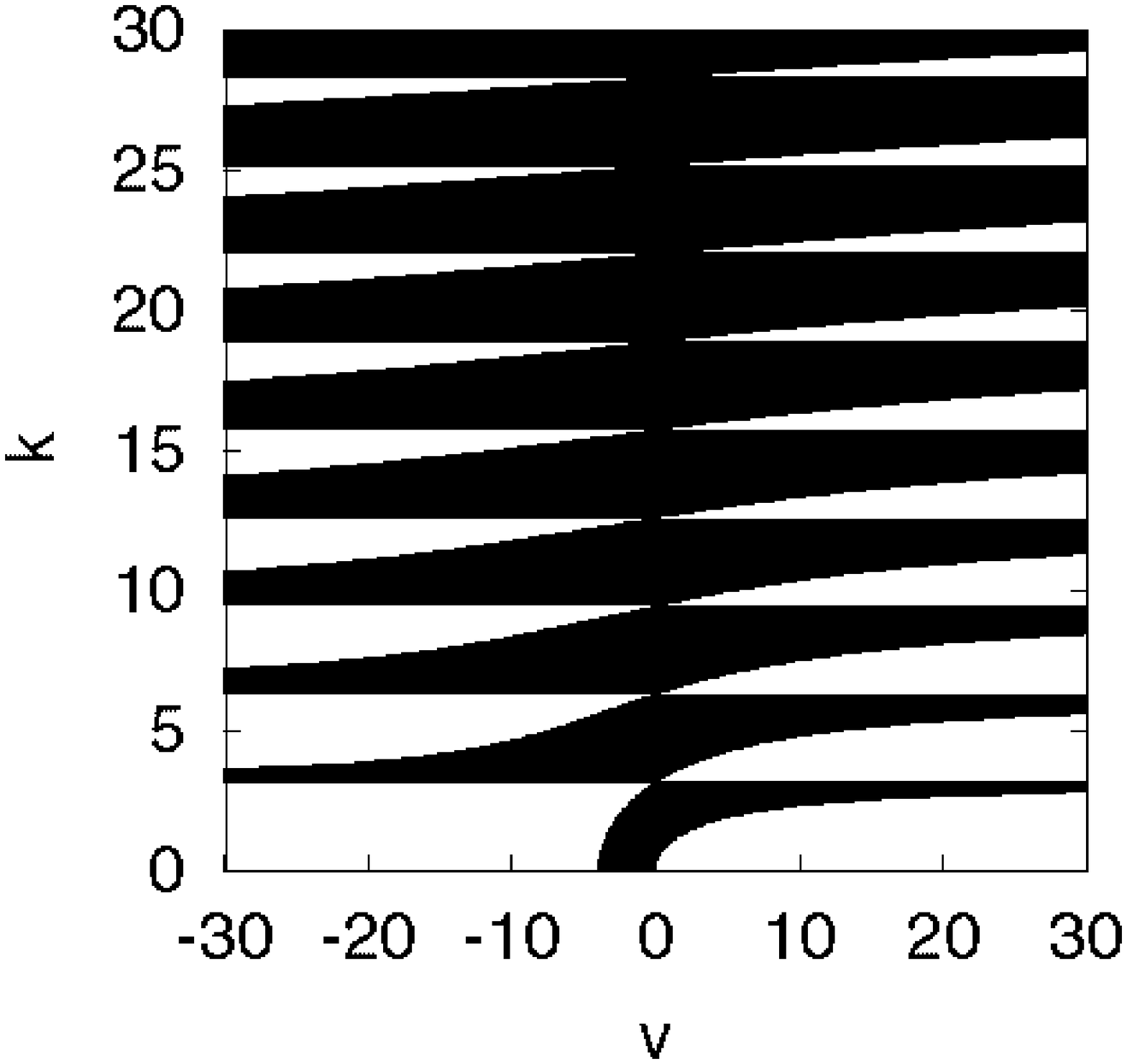}
\includegraphics[width=6.2cm]{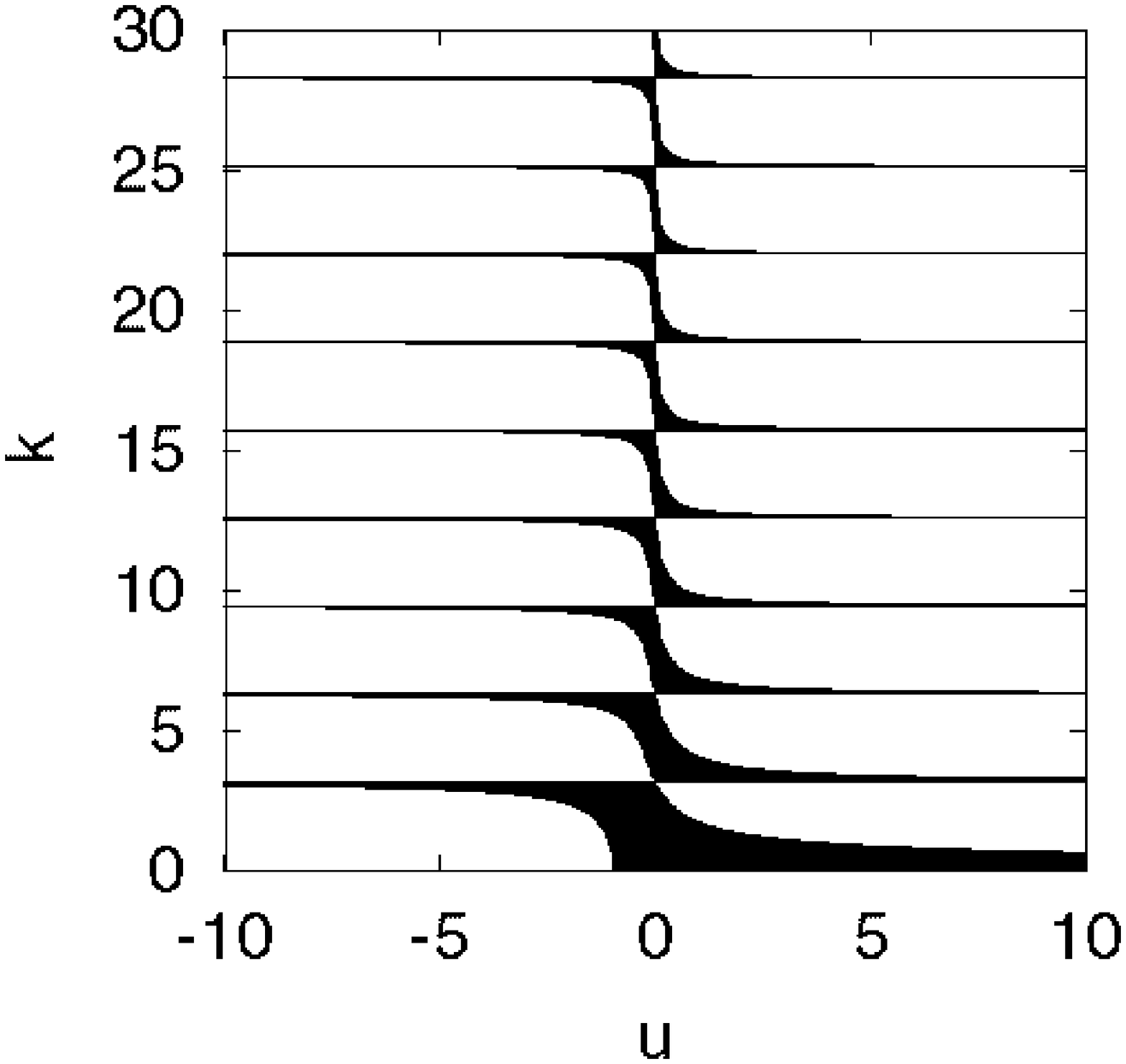}
\includegraphics[width=6.2cm]{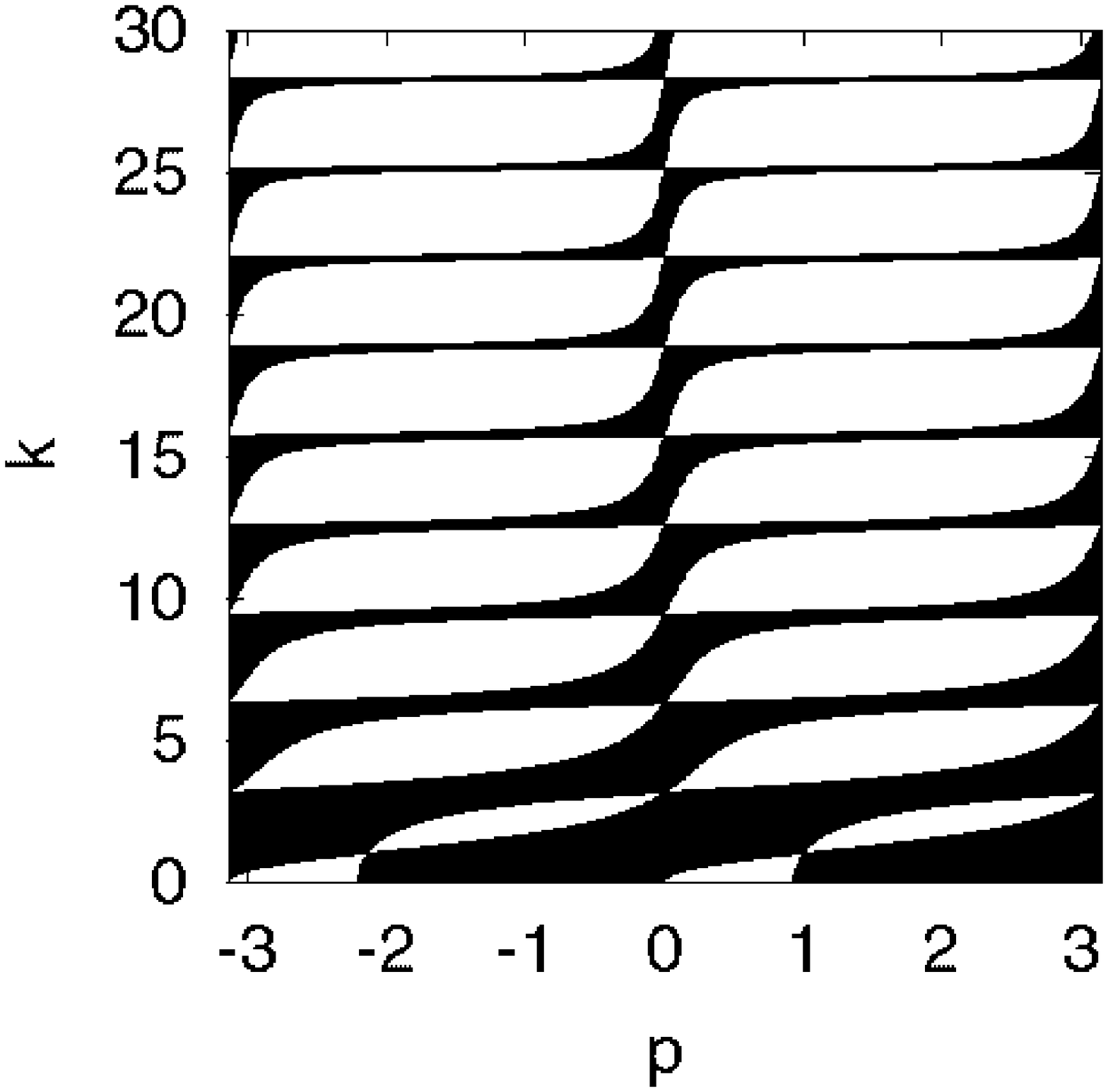}
\includegraphics[width=6.2cm]{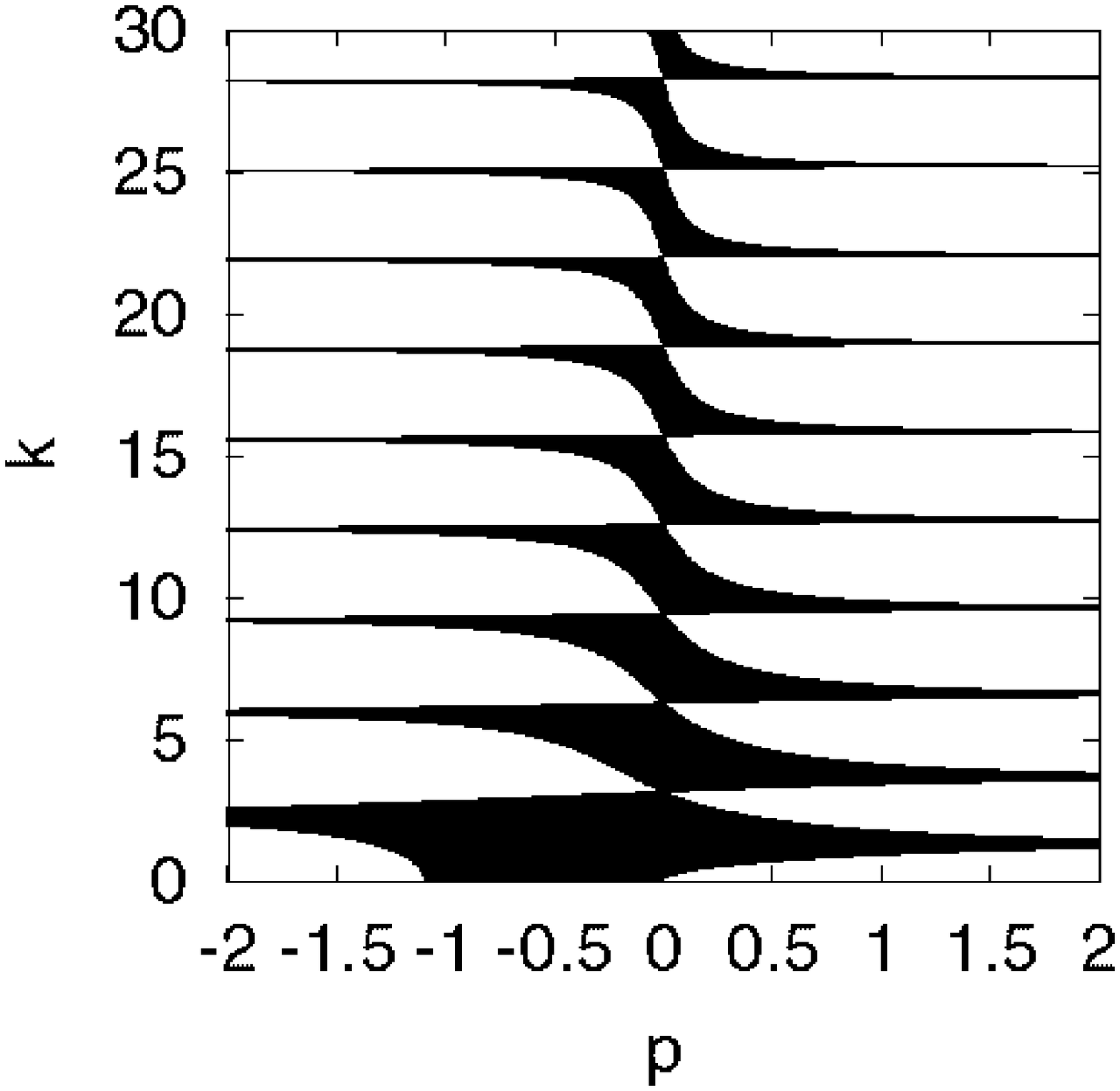}
\caption
{\label{fig3}
The energy dependence of the band spectrum. 
The indications are the same as in Fig.2 except that 
the vertical axis is the wave number $k_0$. 
}
\end{figure}

\vspace{5mm}
\noindent
The reason is easy to understand by examining the 
scattering properties by a single contact interaction. 

In the transfer matrix formulation, 
the bi-orthogonal eigenvectors (\ref{eq2-7}) and (\ref{eq2-9}) 
serve to examine the scattering properties. 
Since $e^{\pm ik_0x}{\bf u}_{\pm k_0}$ 
are the solutions of the equation (\ref{eq2-3}),  
the wave function is written as 
\begin{eqnarray}
\label{eq3-9}
\hspace*{-3ex}
{\bf \Psi}(x)= 
\left\{ \begin{array}{ll}
e^{ik_0x}{\bf u}_{+k_0} + R_S e^{-ik_0x}{\bf u}_{-k_0}, & (x<0), \\
T_S e^{ik_0x}{\bf u}_{+k_0}, & (x>0), \\
\end{array}\right.
\end{eqnarray}
where $T_S$ and $R_S$ are transmission and reflection coefficients, 
respectively. 
We assume that the contact interaction is placed at $x=0$ and 
the incident wave comes from minus infinity in Eq.(\ref{eq3-9}).   
From the connection condition (\ref{eq1-4}), we obtain 
\begin{eqnarray}
\label{eq3-10}
T {\bf u}_{+k_0} = {\cal V} ({\bf u}_{+k_0} + R_S {\bf u}_{-k_0}). 
\end{eqnarray}
Multiplying ${\bf v}_{+k_0}^{\dagger}{\cal V}^{-1}$ from the left and 
using the bi-orthogonal relations (\ref{eq2-10}), 
we can estimate the transition probability as 
\begin{eqnarray}
\label{eq3-11}
\hspace*{-3ex}
|T|^2 & 
\hspace{-1ex}
= &  
\hspace{-1ex}
\left| {\bf v}_{+k_0}^{\dagger} 
{\cal V}^{-1}{\bf u}_{+k_0} \right|^{-2} \nonumber \\
& 
\hspace{-1ex}
= & 
\hspace{-1ex}
4\left[ \alpha^2+\gamma^2+2 +
\delta^2 \frac{k_0^2}{4m^2} + \beta^2 \frac{4m^2}{k_0^2} 
\right]^{-1}. 
\end{eqnarray}
The reflection probability is given by 
\begin{eqnarray}
\label{eq3-12}
|R|^2 = 1- |T|^2. 
\end{eqnarray}
Eq.(\ref{eq3-11}) shows $|T|^2 \longrightarrow 0$ as 
$k_0\longrightarrow +\infty$, 
namely perfect reflection in the high energy limit for generic cases.  
The exception arises in case of $\delta=0$, namely $\delta$ potential. 
In this case,  we have 
\begin{eqnarray}
\label{eq3-13}
|T_{\delta}|^2 & = & 
\left[ 1 + \frac{v^2 m^2}{k_0^2} \right]^{-1}, \\
\label{eq3-14}
|R_{\delta}|^2 & = & 
\left[ 1 + \frac{k_0^2}{v^2 m^2} \right]^{-1}. 
\end{eqnarray}
Thus, $|T_{\delta}|^2 \longrightarrow 1$ as $k_0\longrightarrow +\infty$,   
namely perfect transmission is realized. 
This explains why the band width broadens even in the $k_0$ 
(wave number) space for the $\delta$ array 
as the energy increases and the band gap disappears in 
the high energy limit. 

In the low energy limit,  
the $\varepsilon$ potential rather than $\delta$ 
shows a special nature.  
From Eqs.(\ref{eq3-11}) and (\ref{eq3-12}), 
one recognizes in generic cases, 
$|T|^2 \longrightarrow 0$ as $k_0\longrightarrow 0$, 
i.e. perfect reflection in the low energy limit.  
The exception arises in case of $\beta=0$, 
namely $\varepsilon$ potential. 
In this case, we have 
\begin{eqnarray}
\label{eq3-15}
|T_{\varepsilon}|^2 & = & 
\left[ 1 + \frac{u^2 k_0^2}{16m^2} \right]^{-1}, \\
\label{eq3-16}
|R_{\varepsilon}|^2 & = & 
\left[ 1 + \frac{16m^2} {u^2 k_0^2}\right]^{-1}. 
\end{eqnarray}
Thus, $|T_{\varepsilon}|^2 \longrightarrow 1$ as 
$k_0\longrightarrow 0$, namely perfect transmission 
is realized. 

\section{Conclusion}

We have formulated the generalized Kronig-Penney model 
which has a periodic array of generalized contact interactions 
in one dimension.  
The transfer matrix formalism serves to 
deduce the eigenvalue equation which determines 
the dispersion relation.  

Numerical analysis of the band spectra shows that 
the band structure by the usual periodic $\delta$ array is not 
generic 
and the band width becomes broader in the $k$ space as 
the energy increases, 
while it tends to be narrower in generic cases. 
This fact opens up an interesting possibility that several distinct classes 
of band structures found in various materials may be modeled 
and classified in terms of simple and solvable 
but versatile model of generalized Kronig-Penney hamiltonian.
It would be of interest to extend the current approach
to related problems with direct experimental relevance,
such as the Wannier-Stark ladder problem \cite{AE94, ADE98}.

\vspace*{3ex}

We thank Professor Toshiya Kawai and Professor Kazuo Takayanagi
for valuable discussions and comments.


\begin{thebibliography}{99}
%
\bibitem{AG88}%
S. Albeverio, F. Gesztesy, R. H{\o}egh-Krohn and H. Holden, \\
``Solvable models in quantum mechanics'' ,
Springer, New York, 1988.
%
\bibitem{KP31}%
R. de L. Kronig and W. G. Penney, 
``Quantum mechanics of electrons in crystal lattices'',
Proc. Roy. Soc. (London), 
vol.130A, pp.499--513, 1931. 
%
\bibitem{GK85}%
F. Gesztesy and W. Kirsch, 
``One-dimensional Schr\"{o}dinger operators with 
interactions singular on a discrete set'',
{\em J. Reine Angew. Math.}, vol.362, pp.28--50, 1985. 
%
\bibitem{AD98}%
S. Albeverio, L. Dabrowski and P. Kurasov,
``Symmetries of Schr\"{o}dinger operators with point interactions'',
Lett. Math. Phys., vol.45, pp.33--47, 1998.
%
\bibitem{CF01}%
T. Cheon, T. F{\" u}l{\" o}p and I. Tsutsui,
``Symmetry, duality and anholonomy of point interaction in one dimension'',
Ann. of Phys. (NY) vol. 294, pp.1-23, 2001.
%
\bibitem{TF01}%
I. Tsutsui, T. F{\" u}l{\" o}p and T. Cheon,
``M{\" o}bius structure of the Schr{\" o}dinger operators with point interactions'',
J. math. Phys. vol. 42, pp.5687--5697, 2001.
%
\bibitem{S86b}%
P. \v{S}eba, 
``Some remarks on the $\delta'$-interaction in one dimension'',
Rep. Math. Phys., vol.24, pp.111--120, 1986. 
%
%
\bibitem{CS98a}%
T. Cheon and T. Shigehara, 
``Realizing discontinuous wave functions with renormalized short-range 
potentials'',
Phys. Lett., vol.A243, pp.111--116, 1998.  
%
\bibitem{ENZ01}%
P. Exner, H. Neidhardt, and V. Zagrebnov, 
``Potential approximations to delta': an inverse Klauder phenomenon 
with norm-resolvent convergence'',
Commun. Math. Phys., Vol.224, pp.593--612, 2001. 
%
\bibitem{SM99a}%
 T. Cheon, K. Takayanagi, and T. Shigehara, 
``Equivalence of local and separable realizations of 
the dicontinuity-inducing contact interactionss'',
J. Phys. Soc. Jpn., Vol.69, pp.345--350, 2000. 
%
\bibitem{BBM95}%
R. Balian, D. Bessis, and G. A. Mezincescu, 
``Form of kinetic energy in effective-mass Hamiltonians for heterostructures'',
Phys. Rev. B, vol.51, pp.17624--17629, 1995.  
%
\bibitem{CS99}%
T. Cheon and T. Shigehara, 
``Fermion-boson duality of one-dimensional quantum particles
with generalized contact interactions'',
Phys. Rev. Lett., vol.82, pp.2536--2539, 1999.  
%
\bibitem{C98}%
T. Cheon, 
``Double spiral energy surface in one-dimensional quantum mechanics 
of generalized pointlike potentials'', 
Phys. Lett., vol.A248, pp.285--289, 1998.  
%
\bibitem{AE94}%
J.E. Avron, P. Exner and Y. Last,
``Periodic Schr\"{o}dinger operators with large gaps
and Wannier-Stark ladders'',
Phys. Rev. Lett., vol.72, pp.896-899, 1994.
%
\bibitem{ADE98}%
J. Asch, P. Duclos, P. Exner, 
``Stability of driven systems with growing gaps. 
Quantum rings and Wannier ladders'', 
J. Stat. Phys., vol.92, pp.1053--1069, 1998.  


\end{thebibliography}
\end{document}